\documentstyle{article}
\font\Sets=msbm10
\def\Integer{\hbox{\Sets Z}}
\def\Real{\hbox{\Sets R}}
\def\s#1#2{\langle #1,#2 \rangle}
\def\bb{\begin{equation}}
\def\ee{\end{equation}}
\def\th#1{\noindent {\bf #1} \hspace{1mm}}
\def\proof{\noindent {\bf Proof.} \hspace{1mm}}
\def\qed{\vrule height5pt width4pt depth0pt}

\begin{document}
\title{Integrable deformations of a polygon}
\author{V.E. Adler\\
        Ufa Institute of Mathematics, Russian Academy of Sciences,\\
        Chernyshevsky str. 112, 450000 Ufa, Russia\\
        e-mail: adler@nkc.bashkiria.su}
\date{July 18, 1994}
\maketitle
\thispagestyle{empty}
\vspace{5mm}
\begin{abstract}
The new integrable mapping with a simple geometric interpretation is
presented.  This mapping arise from the nonlinear superposition principle for
the B\"acklund transformations of some vector evolution equation.
\end{abstract}
\vspace{5mm}

\section{Recuttings of the polygon}
Let us consider a set of points $v_j\in\Real^m,\; j\in\Integer_N,\; N\geq 3$
and define the transformation $R_k$ as reflection of the point $v_k$ about
the normal hyperplane passing through the middle of the segment
$v_{k-1}v_{k+1}$ (all other points of the set remain unmovable).
Intuitevely, one cuts off the triangle $v_{k-1}v_kv_{k+1}$ from the polygon
$v_1...v_N$ (that is closed broken line in $\Real^m$), reverses it and glues
it back in the same twodimensional plane.  The formula which defines this
transformation is
\bb R_k:\;\;
 \tilde v_k=v_k+\frac{|v_{k+1}-v_k|^2-|v_k-v_{k-1}|^2}
                     {|v_{k+1}-v_{k-1}|^2}(v_{k+1}-v_{k-1}),\;\;
 \tilde v_j=v_j,\; j\neq k.                             \label{rk}
\ee
Here $|a|^2=\s{a}{a}$ denotes the square of the vector $a$ length, $\s{}{}$
is standard scalar product.  If accidently $v_{k+1}=v_{k-1}$ then $R_k$ is
defined as identity transformation.  Thus some $N$-valued mapping
$R:(\Real^m)^N \rightarrow (\Real^m)^N$  is defined and the problem is to
investigate its iterations.  Transformations $R_1,\dots,R_N$ generate some
group $G$ which acts on $(\Real^m)^N$ and other setting of the problem is to
investigate the dynamics of the vertices $v_j$ under this action.

This elementary geometric model possesses a number of remarkable properties
and demonstrates very regular and beautiful behavior.  In fact, the mapping
$R$ turns out to be an example of integrable mappings, the general theory of
which is actively developed now.  Integrability means existence of the large
enough set of invariants which make possible to apply some discrete version
of the Liouville theorem (see e.g. \cite{ves}, \cite{br}).  The next section
provides a tool for generating of such a set, however the full analysis of
its completeness is very complicated and I do not perform it.  Instead of it
I present some indirect evidences of integrability such as existence of
continuous symmetries of the mapping and polynomial growth of the images
number under its iterations.

The case $m=2$ corresponding to the planar polygons was considered in
\cite{ad}, where connection with the B\"acklund transformation for the KdV
equation was established.  The approach of the present paper is more
straightforward and is suitable for the general case.

In conclusion of this introductory section I mention the most obvious
features of the model.
\begin{enumerate}
\item{It is clear that transformation (\ref{rk}) permutes the quantities
      $\beta_k$ and $\beta_{k-1},$ where $\beta_j=|v_{j+1}-v_j|^2.$
      Therefore the polynomial
      \bb (\lambda+\beta_1)\dots(\lambda+\beta_N)  \label{d0}\ee
      is preserved under action of the group $G$ and provides $N$ invariants.
      Further we shall see that there exist other invariants in addition to
      these.}
\item{The fixed points of the mapping $R$ are obviously all equilateral
      polygons.}
\item{If initial values of all vertices $v_j$ lie on $m'$-dimensional
      hyperplane or sphere in $\Real^m$ then dynamics will be restricted on
      this submanifolds.  In particular we can take without loss of
      generality that $m<N.$}
\item{Other example of invariant manifold come out if $N$ is even and odd
      vertices lie on one and even vertices on the other of two concentric
      spheres or parallel hyperplanes.}
\item{One-dimensional case (all vertices lie on a line or circle) is quite
      trivial and is reduced to the permutations of the polygons sides
      combined with shift or rotation.}
\end{enumerate}

\section{Zero curvature representation}
As in the continuous case the existence of some matrix representation like
L-A pair is very important for studying of mappings.  There exist several
modifications of this notion for the discrete case, see e.g. \cite{ves},
\cite{mv}, \cite{br}.  We shall use the discrete version of the zero
curvature representation presented in \cite{ad}, \cite{ay}.  Let $v_j$ be
column vectors and
\bb  W=\left(
 \begin{array}{ccc}
               -|v_1|^2        &             2v^t_1              &  2  \\
 -(\lambda+\beta)v_1-|v_1|^2v  &   2vv^t_1+(\lambda+\beta)I_m    &  2v \\
  {1\over 2}(\lambda+\beta)(\lambda+|v_1|^2+|v|^2)
      +{1\over 2}|v_1|^2|v|^2  & -(\lambda+\beta)v^t-|v|^2v^t_1  & -|v|^2
 \end{array}\right),
                                                        \label{w}\ee
where $\beta=|v_1-v|^2$ and $I_m$ is $m\times m$ identity matrix.  Let $W_j$
be the matrix $W$ in which $v_1$ and $v$ are substituted by $v_{j+1}$ and
$v_j$ correspondingly.  The following statement is proved directly.

\th{Proposition 1.} The transformation $R_k$ is equivalent to the following
relations:
\bb R_k:\;\; \tilde W_k\tilde W_{k-1} = W_kW_{k-1},\;\;
             \tilde W_j =W_j,\; j \neq k,k-1.           \label{ww}\ee

This representation implies some important consequences.

\th{Corollary 1.} The characteristic polynomial
\bb d(\mu,\lambda) = \mu^{m+2}+d_{m+1}(\lambda)\mu^{m+1}
     +\dots d_0(\lambda) =\mbox{det}(\mu I_{m+2}-\hat W_1)  \label{inv}\ee
where $\hat W_1=W_N\dots W_2W_1$ is invariant under action of the
transformations $R_j.$

\proof
Obviously the transformations $R_N,\dots,R_2$ do not change $\hat W_1.$  The
transformation $R_1$ acts as conjugation:
$$ \tilde W_N\dots W_2\tilde W_1=
 (\tilde W_NW^{-1}_N)\hat W_1(\tilde W_NW^{-1}_N)^{-1} $$
and therefore does not change characteristic polynomial as well.
\qed

It is easy to check that $\mbox{det}\,W = -(\lambda+\beta)^{n+2}$ so that
invariant (\ref{d0}) is equivalent to $d_0.$  The structure of the other
terms in (\ref{inv}) is not clear to me now.  Apparently this invariants do
not exhaust all invariants of the mapping.  It seems that $d(\mu,\lambda)$ is
polynomial on the values $|v_i-v_j|^2$ and therefore it is invariant under
shift $v\rightarrow v+a.$  On the other hand, the mapping $R$ {\em admits}
invariants which are not invariant under the shift and therefore are
functionally independent with $d(\mu,\lambda).$  An example of such invariant
is vector
 \bb J = \sum^N_1|v_j|^2(v_{j+1} - v_{j-1}).            \label{j}\ee
Applying the shift we obtain from it the new invariant
 \bb S = \sum^N_1\s{a}{v_j}(v_{j+1}-v_{j-1}),           \label{s}\ee
where $a$ is arbitrary constant vector in $\Real^m.$  Obviously $S$ is
orthogonal to $a,$ so both vectors $J$ and $S$ provide $2m-1$ scalar
invariants.  The invariants $S$ corresponding to the different vectors $a$
are equivalent under orthogonal group $SO(m).$

\th{Corollary 2.} Transformations (\ref{rk}) satisfy identities
\bb R^2_j= (R_jR_{j+1})^3= (R_iR_j)^2= 1, \; i\neq j\pm 1.
                                                        \label{id}\ee
\proof
We have to prove only that $(R_jR_{j+1})^3=1$ since other identities are
rather obvious.  Denote $Q=(R_jR_{j+1})^3$ and $Q(W_i)=\hat W_i.$  Consider
the product $P=W_{j+1}W_jW_{j-1}.$  Accordingly to the proposition 1,
transformation $Q$ does not change it, that is $\hat P=P.$  Moreover, it is
clear that $Q$ acts identically on the determinants of the matrices $W_i.$
Assume $\lambda=\beta_{j+1}$ then the first factors in both products $P$ and
$\hat P$ become degenerate.  The image of the matrix $W_{j+1}$ is spanned
over $m+2$-vector $(2,2v^t_{j+1},-|v_{j+1}|^2)^t$ and since the matrix
$W_jW_{j-1}$ is nondegenerate, it coincides with the image of $P.$
Analogously, the images of $\hat W_{j+1}$ and $\hat P=P$ coincide and
therefore $\hat v_{j+1}=v_{j+1}.$  The cancellation of the common factor
$W_{j+1}$ implies equality $\hat W_j\hat W_{j-1}=W_jW_{j-1}$ and repetition
of the argument proves the statement.
\qed

This means that the group $R$ is isomorphic to the affine Weyl group $\tilde
A_{N-1},$ or in other words, transformations $R_j$ define the nonlinear
representation of the $\tilde A_{N-1}.$

\th{Corollary 3.} If one performs recuttings $R_j$ of the polygon excepting
one of them, say $R_1,$ then all images of the points $v_j$ form finite set.

\proof
The subgroup generated by $R_j,\;j\neq 1$ is isomorphic to the symmetric
group $S_N$ and therefore is finite.
\qed

For the general $N$-valued mapping the number of initial data images under
iterations grows exponentially.  Accordingly to \cite{ves} one of the
integrable mappings characteristic features is the polynomial growth of the
images number.  Using the identities (\ref{id}) one can prove that this
criterion is satisfied in our case.  In fact, the transformations
$$ T_j= (R_{j+N-1}\dots R_{j+1}R_j)^{N(N-1)},\; j=1,\dots,N-1 $$
generate the commutative subgroup with finite index in $G$ and therefore
the number of images after $k$ iterations grows as $Ck^{N-1}.$

\section{Continuous motions of the polygon}
Some continuous symmetries of the presented mapping are quite obvious from
its geometric description.  Really, any shift or rotation of $\Real^m$
commutes with transformations $R_j$ as well as reflections and homotheties.
We have already used this fact for generating invariant (\ref{s}) from the
invariant (\ref{j}).  Analogy between the transformations (\ref{rk}) and
examples considered in \cite{ad}, \cite{ay} suggests that there exist some
hidden symmetries as well.  Now we shall demonstrate that it is really so.
Let us consider the lattice
\bb |v_{j+1}-v_j|^2 (v_{j+1}+v_j)_x
              = 2\s{v_{j,x}}{v_{j+1}-v_j}(v_{j+1}-v_j). \label{vx}\ee
Scalar product of (\ref{vx}) with $v_{j+1}-v_j$ yields
\bb \s{v_{j+1,x}-v_{j,x}}{v_{j+1}-v_j}= 0,              \label{z1}\ee
that is the values $\beta_j$ are the first integrals of the lattice.  Then
product with $v_{j+1,x}-v_{j,x}$ yields
$$ |v_{j+1,x}|^2 = |v_{j,x}|^2,                         $$
that is the velocities of all the points are of the same absolute value.  We
assume the normalization
\bb  |v_{j,x}|^2=1,\;j=1,\dots,N                        \label{norm}\ee
without loss of generality since the lattice (\ref{vx}) is underdetermined
and is invariant under change of independent variable $\tilde x= \phi(x)$
with arbitrary $\phi.$

The lattice (\ref{vx}) admits the zero curvature representation
\bb   W_{j,x} = U_{j+1}W_j - W_jU_j                     \label{wu}\ee
where $W_j$ is given by (\ref{w}) and the matrix $U_j$ is defined by formula
\bb  U= {1\over\lambda}\left(
  \begin{array}{ccc}
        -2\s{v}{v_x}     &         2v^t_x             &   0        \\
  |v|^2v_x-2\s{v}{v_x}v  &     2vv^t_x-2v_xv^t        &  -2v_x     \\
             0           & 2\s{v}{v_x}v^t-|v|^2v^t_x  & 2\s{v}{v_x}
  \end{array}\right)                                    \label{u}\ee
where $v$ is substituted by $v_j.$

\th{Corollary 4.} The characteristic polynomial (\ref{inv}) is the first
integral of the lattice (\ref{vx}).

\proof
This is obvious from the Lax equation $\hat W_{1,x}=[U_1,\hat W_1]$ which
follows from (\ref{wu}).
\qed

For the examples considered in \cite{ay} the consistency of the
transformations (\ref{rk}) and lattice (\ref{wu}) is proved by arguments
which are based on special structure of the matrices $W$ and $U$ and fail in
our case (the obstacle is the presence of derivatives $v_x$ in $U.$)
Nevertheless, the following proposition can be proved by direct, although
tedious computations.

\th{Proposition 2.} Transformations $R_j$ act on the system (\ref{vx}),
(\ref{norm}).

The representation (\ref{wu}) suggests that the lattice (\ref{vx}) can be
regarded as sequence of the B\"acklund transformation (BT) for some partial
differential equation with zero curvature representation
\bb  U_t = V_x - [V,U].                                 \label{uv}\ee
The search of the matrix $V$ brings us to
$$ V= ({3\over 2}\s{v_{xx}}{v_{xx}}-{4\over\lambda})U +{2\over\lambda}V_0 $$
where
$$  V_0= \left(
 \begin{array}{ccc}
  -\s{v}{v_{xxx}}  &  v^t_{xxx}  &  0  \\
   u  &  vv^t_{xxx}-v_{xxx}v^t+2v_{xx}v^t_x-2v_xv^t_{xx} & -v_{xxx} \\
            0      &    -u^t     &  \s{v}{v_{xxx}}
  \end{array}\right)  $$
and
$u = {1\over 2}|v|^2v_{xxx} - \s{v}{v_{xxx}}v +
      2\s{v}{v_{xx}}v_x - 2\s{v}{v_x}v_{xx} + 2v_x.$

The corresponding equation (\ref{uv}) is equivalent to
\bb  v_t = v_{xxx}+{3\over 2}\s{v_{xx}}{v_{xx}}v_x,\;
     \s{v_x}{v_x}=1.                                    \label{vt}\ee

The folowing statement can be proved directly.

\th{Proposition 3.} The flows defined by the lattice (\ref{vx}),(\ref{norm})
and equation (\ref{vt}) commute.  Transformations (\ref{rk}) are consistent
with the last equation.

One can easily check that the formula (\ref{rk}) can be rewritten as
$$  v_{k+1}=v_{k-1}+\frac{|\tilde v_k-v_{k-1}|^2-|v_{k-1}-v_k|^2}
                         {|\tilde v_k-v_k|^2}(\tilde v_k-v_k)     $$
(this is obvious also from geometric interpretation).  In this form it can be
considered as nonlinear superposition formula which allows one to construct
the solution $v_{k+1}$ of the equation (\ref{vt}) connected with the solution
$v_{k-1}$ by double BT by use of two solutions $v_k$ and $\tilde v_k$ which
are results of the single BT.

The equation (\ref{vt}) appeared in \cite{ss} for the first time.  In the
case $m=1$ the lattice (\ref{vx}),(\ref{norm}) and equation (\ref{vt}) are
trivial as well as the mapping $R.$  In the case $m=2$ denote $v=(p,q)^t,$
then the equation (\ref{vt}) is reduced to the scalar equation
$$ p_t = p_{xxx} - {3\over 2}{p_xp^2_{xx}\over p^2_x-1} $$
which is well known to be integrable.

\section{Concluding remarks}
Some problems concerning the presented models were disregarded.  First of
all, their Hamiltonian structures should be investigated.  Possibly some
analogies with the case of the dressing chain \cite{vsh} will be useful here.

Other problem is to study higher symmetries of the equation (\ref{vt}).  I
hope that recursion operator can be found and the whole hierarchy
constructed.

Recently a number of integrable equations was shown to describe motion of the
nonstretching curves, see e.g. \cite{gp} and \cite{ds} where the condition
that curve lies on $m$-dimensional sphere naturally arises.  Possibly the
presented model and its continuous symmetries can be considered as
quantization of such motion.

Of course we can assume that number of points is infinite so that
$j\in\Integer.$  This corresponds to recuttings of nonclosed broken line.  I
do not know if there exists some other boundary condition besides the
periodic closure which leads to integrability of the mapping $R$ and the
lattice (\ref{vx}).  Possibly some modification of the approach from
\cite{vsh}, \cite{ad} can bring to the Painlev\'e type equations and their
B\"acklund transformations.

\section{Acknowledgements}
I should like to thank S.I. Svinolupov for useful discussions.  This work was
partially supported by International Science Foundation.



\begin{thebibliography}{99}
\bibitem{ves}{A.P. Veselov (1991) Integrable mappings, Uspekhi Mat. Nauk
             46(5), 3-45}
\bibitem{br}{M. Bruschi, O. Ragnisco, P.M. Santini, Tu Gui Zhang (1991)
             Integrable symplectic maps, Physica D 49, 273-294}
\bibitem{ad}{V.E. Adler (1993) Recuttings of polygons, Funkts. Analiz
             27(2), 79-82}
\bibitem{mv}{J. Moser and A.P. Veselov (1991) Discrete versions of some
             classical integrable systems and factorization of matrix
             polynomials, Commun. Math. Phys. 139, 217-243}
\bibitem{ay}{V.E. Adler and R.I. Yamilov (1994) Explicit auto-transformations
             of integrable chains, J. Physics A 27, 477-492}
\bibitem{gp}{R.E. Goldstein and D.M. Petrich (1991) The Korteweg -- de Vries
             hierarchy as dynamics of closed curves in the plane, Phys. Rev.
             Lett. 67(23), 3203-3206}
\bibitem{ds}{A. Doliwa and P.M. Santini (1993) An elementary geometric
             characterization of the integrable motions of a curve, Preprint
             IFT/11/93}
\bibitem{vsh}{A.P. Veselov, A.B. Shabat (1993) Dressing chain and spectral
             theory of the Schr\"odinger operator, Funkts.  Analiz 27(2),
             1-21}
\bibitem{ss}{V.V. Sokolov, S.I. Svinolupov (1994) Vector and matrix
             generalizations of the integrable equations of
             mathematical physics, Teoret. i Mat. Fiz., to appear}
\end{thebibliography}
\end{document}